\documentstyle[12pt,epsf]{article}
\def\gsim{\;\raise0.3ex\hbox{$>$\kern-0.75em\raise-1.1ex\hbox{$\sim$}}\;}
\def\lsim{\;\raise0.3ex\hbox{$<$\kern-0.75em\raise-1.1ex\hbox{$\sim$}}\;}
\newcommand{\be}{\begin{equation}}

\newcommand{\ee}{\end{equation}}
\newcommand{\bea}{\begin{eqnarray}}
\newcommand{\eea}{\end{eqnarray}}
\newcommand{\bt}{\begin{tabular}}
\newcommand{\et}{\end{tabular}}
\newcommand{\ba}{\begin{array}}
\newcommand{\ea}{\end{array}}

\newcommand{\bvec}{\mathbf}

\begin{document}

\thispagestyle{empty}

\setcounter{page}{0}

{}\hfill{DSF$-$6/2004}

{}\hfill{physics/0406030}

\vspace{1truecm}

\begin{center}
{\Large \bf Fermi, Majorana and the statistical model of atoms}
\end{center}

\bigskip\bigskip

\begin{center}
{\bf E. Di Grezia$^{1,2,a}$ and S. Esposito$^{1,2,3,b}$}

\vspace{.5cm}

$^1$ {\it Dipartimento di Scienze Fisiche, Universit\`{a} di
Napoli ``Federico II'' \\ Complesso Universitario di Monte S.
Angelo, Via Cinthia, I-80126 Napoli, Italy}

$^2$ {\it Istituto Nazionale di Fisica Nucleare, Sezione di
Napoli, Complesso Universitario di Monte S. Angelo, Via Cinthia,
I-80126 Napoli, Italy}

$^3$ {\it Unit\`{a} di Storia della Fisica, Facolt\`{a} di Ingegneria,
Universit\`{a} Statale di Bergamo, Viale Marconi 5, I-24044 Dalmine
(BG), Italy}

$^a$ e-mail address: Elisabetta.Digrezia@na.infn.it

$^b$ e-mail address: Salvatore.Esposito@na.infn.it
\end{center}

\bigskip\bigskip\bigskip

\vspace{3cm}
\begin{abstract}
\noindent We give an account of the appearance and first
developments of the statistical model of atoms proposed by Thomas
and Fermi, focusing on the main results achieved by Fermi and his
group in Rome. Particular attention is addressed to the unknown
contribution to this subject by Majorana, anticipating some
important results reached later by leading physicists.
\end{abstract}

${}$ \\

%\noindent Keywords:  Majorana and Quantum Mechanics, Weyl
%reformulation of Quantum Mechanics, Group Theory and Quantum
%Mechanics, Symmetries in Physics

\newpage

\section{Introduction}
In the occasion of the centennial of the death of A. Volta, an
important international conference was held in Como (Italy) on
September 1927, mainly devoted to the principles and applications
of the new Quantum Mechanics. Among the interesting contributions
by Bohr, Heisenberg, Pauli and others, it stands out the long
speech by Fermi aimed to point out the different behavior of the
particles satisfying the Bose-Einstein statistics and those
obeying the Pauli exclusion principle, such as electrons. Fermi
started to investigate in this direction since 1923 \cite{[FN16]},
\cite{[FN19]}, when he published his ``remarks on the quantization
of systems with identical particles", where anticipated the
formulation of the exclusion principle for generic molecules given
by Pauli one year later. Soon after the appearance of the Pauli
paper, Fermi recognized to have all the tools in order to build a
theory of perfect gases satisfying the Nernst principle at zero
temperature, and on the 7th February, 1926 he presented the
statistical distribution law for particles obeying the exclusion
principle \cite{[FN30]}. Such a law will be independently
discovered later by Dirac in the August of the same year. The most
distinguished physicists of that epoch reached immediately the
importance of the Fermi work, and in the February of 1927 Pauli
applied it to the conduction electrons in a metal, explaining
their anomalous paramagnetism. Later on, at the Como conference,
Sommerfeld reported on his recent works on thermal and transport
properties in metals and succeeded in explaining, for the first
time, the electron contribution to the specific heat of a metal.
Probably encouraged by the successful application to specific heat
and entropy calculations, in the fall of 1927, just after the Como
conference, Fermi matured the idea of applying his statistical
method to the completely degenerate state of the electrons in an
atom, in order to evaluate the effective potential acting on the
electrons themselves. He was aware of the fact that, since the
number of the electrons involved is usually only moderately large,
the results obtained will not present a very high accuracy.
Nevertheless, the method resulted to be very simple, and gave an
easy-to-use expression for the screening of the Coulomb potential
accounted for by electrons as a whole.

Then Fermi started to apply his method to several problems of
atomic physics, and some more were suggested at that time to
Rasetti and other associates of his group in Rome. The basic idea
of the statistical method of atoms became one of his preferred
ones, and practically the main activity of the Fermi group at the
Physics Institute in Rome from 1928 to 1932 was devoted to this
subject.

\section{The Thomas-Fermi model}
\subsection{The case of neutral atoms}
The main idea of the statistical model of atoms is that of
considering the electrons around the atomic nucleus as a gas of
particles, obeying the Pauli exclusion principle, at the absolute
zero of temperature. The limiting case of the Fermi statistics for
strong degeneracy applies to such a gas. Then, the maximum
electron kinetic energy (in a neutral atom),
\begin{equation}
\frac{p_0^2}{2m} = e\,V , \label{1}
\end{equation}
can be identified with that of a uniform gas of electrons whose
number density is given by:
\begin{equation}
n= \frac{8\pi p_0^3}{3h^3}  , \label{2}
\end{equation}
Note that the total energy has to be constant at any spatial point
since, otherwise, a flux of electrons from one point to another
would be established. The potential energy $-eV$ depends, thus, on
the position through the electron charge density $\rho = -e n$ at
that point:
\begin{equation}
\rho =
-\frac{1}{3\pi^2}\left(\frac{2\,m}{\hbar^2}\right)^{3/2}(e\,V)^{3/2}
. \label{3}
\end{equation}
This was realized by Thomas in 1926 \cite{[Thomas]} and,
independently, by Fermi in the December of 1927 \cite{[FN43]}.
Fermi and many other relevant physicists\footnote{With the
probable exception of Bohr and Kramers, whose encouragement to
Thomas was acknowledged by Thomas himself at the end of his paper.
However, it sounds strange that at the Como Conference Bohr or
others did not cite the work by Thomas.} were unaware that
essentially identical conclusions had previously been reached by
Thomas, since the English researcher published his results on a
journal which was probably not widespread through the physics
community. According to Rasetti \cite{[FNM]}, ``Fermi became
acquainted with Thomas' paper only late in 1928, when it was
pointed out to him by one (now unidentified) of the foreign
theoreticians visiting Rome". It is also intriguing to note that
for some time the statistical model was denoted as the
Fermi-Thomas model rather then as the Thomas-Fermi one (compare,
for example the papers in Refs \cite{[Baker]},\cite{[Peierls]}
appeared in 1930 with that of Sommerfeld in \cite{[SommZeit]} of
1932). We further observe that Thomas made use of only the
exclusion principle (in writing Eq. (\ref{2}), while Fermi
considered also the dependence of the effective potential
(entering in Eq. (\ref{3})) on the temperature of the gas. Such a
difference, however, is not very significative since the number
density of electrons in the occupied states of an atom reaches
high values due to the small space region where they are placed.
Then, the electrons form a completely degenerate gas, and the
correction to the effective potential reduces to a small term
proportional to the temperature $T$, which does not alter the main
results of the model in the limit $T \rightarrow 0$.

The effective electrostatic potential is related to the electron charge
density by means of the Poisson equation,
\begin{equation}
\nabla^2 V = -4\pi\rho , \label{4}
\end{equation}
and substituting in this the expression in Eq. (\ref{3}), Thomas
and Fermi found a second-order inhomogeneous differential equation
for $V$ with a term proportional to $V^{3/2}$:
\begin{equation}
\nabla^2 V =
\frac{4e}{3\pi}\left(\frac{2\,m}{\hbar^2}\right)^{3/2}(e\,V)^{3/2}
. \label{5}
\end{equation}
This equation allows the evaluation of the potential inside an
atom with atomic number $Z$, using boundary conditions such that
for the radius $r\rightarrow 0$ the potential becomes the Coulomb
field of the nucleus,
\begin{equation}
 V(r) \rightarrow\frac{Ze}{r}  ,
\label{6}
\end{equation}
while for $r \rightarrow \infty$:
\begin{equation}
 V(r) \rightarrow 0  .
\label{7}
\end{equation}
The additional condition for the total charge,
\begin{equation}
\int n \, d^3{\bvec{r}}= Z  , \label{8}
\end{equation}
is automatically satisfied, provided that Eqs. (\ref{6}) and
(\ref{7}) hold.

In order to simplify the differential equation in Eq. (\ref{5}),
Fermi introduced a suitable change of variables:
\begin{equation}
\begin{array}{rcl}
r &=& \mu x
\\ & & \\
V(r) &=& \displaystyle \frac{Ze}{r}\varphi (r)
\end{array}
\label{9}
\end{equation}
with
\begin{equation}
\mu = \frac{1}{2}\left(\frac{3\pi}{4}\right)^{2/3}\frac{\hbar^2}{m
e^2}Z^{-1/3}  . \label{10}
\end{equation}
In terms of the new variables, Eq. (\ref{5}) becomes (for $\varphi
> 0$):
\begin{equation}
\varphi^{\prime\prime} = \frac{\varphi^{3/2}}{\sqrt{x}} \label{11}
\end{equation}
(a prime denotes differentiation  with respect to $x$) with the
boundary conditions:
\begin{equation}
\begin{array}{rcl}
\varphi (0) & = & 1,
\\ & & \\
\varphi (\infty) & = & 0.
\end{array}
\label{12}
\end{equation}
The Fermi equation (\ref{11}) is a universal equation which does
not depend neither on $Z$ nor on physical constants ($\hbar, m ,
e$). Its solution gives, from Eq. (\ref{9}), as noted by Fermi
himself, a screened Coulomb potential which at any point is equal
to that produced by an effective charge
\begin{equation}
Z e \, \varphi\left(\frac{r}{\mu}\right). \label{13}
\end{equation}
As was immediately realized, in force of the independence of Eq.
(\ref{11}) on $Z$, the method gives an effective potential which
can be easily adapted to describe any atom with a suitable scaling
factor, according to Eq. (\ref{13}).

\subsection{The case of positive ions}
In order to evaluate ionization energies and similar quantities
which are relevant for Atomic Physics observations, Fermi and
Rasetti \cite{[FNM]}, \cite{[FN45]}, \cite{[FN49]} proceeded to
apply and enlarge the above method to describe positive ions. They
considered the ion of nuclear charge $Ze$ as a neutral atom of
nuclear charge $(Z-1) e$, to be treated with the statistical
method, plus an extra proton in the nucleus.

Since the electrostatic potential for the atom of atomic number
$Z-1$ is, from (\ref{13}),
\begin{equation}
\frac{(Z-1) e}{r} \, \varphi\left(\frac{r}{\mu}\right). \label{14}
\end{equation}
(with $Z\rightarrow Z-1$ for the parameter $\mu$ in Eq.
(\ref{10})), the potential energy of one extra electron will be,
at a first approximation, given by:
\begin{equation}
-eV = -  \frac{e^2}{r} \left[1 + (Z-1) \,
\varphi\left(\frac{r}{\mu}\right)\right] . \label{15}
\end{equation}
With this formula, Fermi and his associates in Rome extended their
calculations to many physical problems, obtaining quantities which
well fitted the observations and the knowledge of the time.

The method was further generalized for arbitrary positively
charged ions by Baker in June 1930 \cite{[Baker]} and corrected
(for an interpretational misunderstanding) by Guth and Peierls six
months later \cite{[Peierls]}.

\section{Solution of an equation}
\subsection{Numerical works}
The problem of the theoretical calculation of observable atomic
properties is solved, in the statistical model approximation, in
terms of the function $\varphi (x) $ introduced in Eq. (\ref{9})
and satisfying the Fermi differential equation (\ref{11}).
However, it was believed that the solution of such equation
satisfying the appropriate boundary conditions in (\ref{12})
couldn't be expressed in closed form, so that some effort was made
by Thomas \cite{[Thomas]}, Fermi \cite{[FN43]},  \cite{[FN49]} and
others \cite{[Baker]} in order to achieve numerical integration of
the differential equation.

Thomas used a numerical method described by Whittaker and Robinson
\cite{[Whittener]} in order to solve the second-order differential
equation for the electrostatic potential $V$. He thus obtained a
numerical table for some mathematical quantities from which one
can deduce the values of $V$ as function of the distance $r$ from
the nucleus. However, as already observed by Baker \cite{[Baker]},
Thomas' numerical calculations of $V$ near $r=0$ are ``slightly in
error", and this influenced also some calculations by Milne
\cite{[Milne]} who directly applied the Thomas theory as early as
in 1927.

A similar effort was also pursued by Fermi who built a numerical
table for the values of the function $\varphi (x)$ obeying Eq.
(\ref{11}). The numerical work was performed in approximately one
week and, according to many testimonies \cite{[FNM]},
\cite{[Amaldi]} (see the anecdote related to Majorana), the table
was ready as early as at the end of 1927, although it was
published only in the German paper of 1928 \cite{[FN49]}. The
numerical values obtained by Fermi were largely used not only by
the members of the Rome group, but even by many other physicists
who dealt with atomic problems (see, for example, Refs.
\cite{[Baker]},\cite{[Sommerfeld]}).

\subsection{Approximate solutions in two limiting cases}
By using standard but involved mathematical tools, in his paper
\cite{[Thomas]} Thomas got an exact, ``singular" solution of his
differential equation satisfying only the condition (\ref{7}).
This was later (in 1930) considered by Sommerfeld
\cite{[Sommerfeld]} as an approximation of the function $\varphi
(x)$ for large $x$ (and is indeed known as the ``Sommerfeld
solution" of the Fermi equation),
\begin{equation}
\varphi (x)=   \frac{144}{x^3} ,
\label{16}
\end{equation}
and Sommerfeld himself obtained corrections to the above quantity
in order to approximate in a better way the function $\varphi (x)$
for not extremely large values of $x$. The goodness of the
approximated expression was checked by Sommerfeld comparing his
results with the numerical values obtained by Fermi
\cite{[Sommerfeld]}.

It is intriguing to observe that the solution in Eq. (\ref{16})
was already known to Majorana \cite{[Esposito]} at the end of 1927
(independently from Thomas), who recognized its crucial role in
determining the desired solution of Eq. (\ref{11}) with boundary
conditions Eq. (\ref{12}) (see below).

An approximate solution for $\varphi (x)$ near $x=0$ was, instead,
first considered by Fermi \cite{[FN43]}, who obtained a series
expansion for it. The reasoning is, probably, as follows. For $x$
near $0$ we have $\varphi (x)\sim 1$, so that substitution into
Eq. (\ref{11}) results in
\begin{equation}
\varphi^{\prime\prime} (x)=   x^{-1/2} . \label{17}
\end{equation}
By integrating this approximate equation we easily have
\begin{equation}
\varphi (x)\sim k_1 + k_2 x  + \frac{4}{3}x^{3/2} ,
\label{18}
\end{equation}
for $x$ near the origin. The constant $k_1$ is obtained directly
from the first condition in (\ref{12}), while $k_2$ was determined
by Fermi to assume the numerical value $-1,58$ for the neutral
atom. Thus, the Fermi approximation for $x\rightarrow 0$ results
to be:
\begin{equation}
\varphi (x)= 1 - 1.58 x  + \frac{4}{3}x^{3/2}+... \, . \label{19}
\end{equation}
According to Rasetti \cite{[FNM]}, Segr\`{e} \cite{[Segre]} and
Amaldi \cite{[Amaldi]}, the paper in Ref. \cite{[FN43]} was shown
to Majorana probably even before its publication. Majorana then
proceeded to improve the degree of approximation of the Fermi
formula (\ref{19}), and considered several other terms in the
series expansion up to the sixth power of $x$ (including both
integer and half-integer powers)\footnote{Majorana never published
his results on the Thomas-Fermi equation. What we discuss here and
in the next section is entirely deduced from the unpublished notes
kept at the Domus Galilaeana in Pisa (Italy) and known, in
Italian, as ``Fogli sparsi" and ``Volumetti" \cite{[Volumetti]}.}.
The method followed by him is a generalization of the Frobenius
method for differential equations \cite{[Frobenius]}; its sketch
is as follows.

Let us consider a solution of Eq. (\ref{11}), which can be rewritten as
\begin{equation}
x(\varphi^{\prime \prime})^2 - \varphi^3 =0  , \label{20}
\end{equation}
in the form
\begin{equation}
\varphi (x)=\sum_{n=0}^\infty (a_{2n} \, x^n + a_{2n - 1} \,
x^{n+\frac{1}{2}}) . \label{21}
\end{equation}
The terms in Eq. (\ref{21}) with integer powers of $x$ account for
the usual Taylor series expansion, while those with half-integer
powers attain to the Frobenius one. The need for both terms is
dictated by the Fermi approximation underlying Eqs.
(\ref{17})-(\ref{19}). The yet unknown coefficients $a_{2n}$ and
$a_{2n - 1}$ are then determined by substituting Eq. (\ref{21})
into Eq. (\ref{20}) and requiring that the coefficients of given
powers of $x$, appearing in the L.H.S. of  Eq. (\ref{20}), vanish.
In such a way, one is able to obtain two iterative formulae for
$a_{2n}$ and $a_{2n - 1}$ coefficients. However, differently from
what is usually found in the notebooks by Majorana (see
\cite{[Volumetti]}), he did not succeed to obtain general
expressions for the coefficients, but he stopped the series to the
term $x^6$ (thus considering a polynomial) and evaluated only some
coefficients. The corresponding expression looks like as follow:
\begin{eqnarray}
\varphi (x) & = & 1 - p x + \frac{4}{3} x^{3/2} -  \frac{2p}{5}
x^{5/2} + \frac{1}{3} x^{3} + \frac{3p^2}{70} x^{7/2}  -
\frac{2p}{15} x^{4} + \frac{56 + 3p^3}{756} x^{9/2} +
\frac{p^2}{175} x^{5} \nonumber \\
&+& \frac{-992p +45p^4}{47520} x^{11/2} + \frac{4(35
-9p^3)}{14175} x^{6} ,
\label{22}
\end{eqnarray}
with $p=1.58$, according to the Fermi value. Probably, Majorana
abandoned the complete series expansion in (\ref{21}) because he
realized that such a series does not converge to the desired
solution. Note, in fact, that the differential equation (\ref{11})
or (\ref{20}) is a non-linear equation, so that the series
solution method cannot, in general, be applied at all. It is
however remarkable that, stopping the series in (\ref{21}) to any
power $n \leq 5$, the obtained polynomial diverges towards to
$+\infty$ for diverging $x$, while from $n=6$ (as considered by
Majorana) onward the corresponding polynomial diverges towards
$-\infty$, as can be seen in Fig.1. Both behaviors, of course, do
not match the correct condition $\varphi(x\rightarrow
\infty)\rightarrow 0$, but we point out that the first one,
$\varphi(x\rightarrow \infty)\rightarrow +\infty$, is physically
not acceptable (see, for example, the discussion in Ref.
\cite{[Peierls]}). Eventually, we point out that the Fermi
approximation of the function $\varphi (x)$ near $x=0$ with a
polynomial was later reconsidered by Baker \cite{[Baker]}, who
obtained terms up to the power $x^{9/2}$. However, the coefficient
of the last term presented by author is wrong (compare with the
correct Majorana result in Eq. (\ref{22})). We have dwell on this
point in order to remark the complexity of calculations, leading
to the expressions for the coefficients in the polynomial
expansion, due to the non-linearity of the equation considered.
For example, in order to evaluate the coefficients $a_i$ in Eq.
(\ref{21}) up to the term $x^{9/2}$, we have to solve a set of
$10$ coupled algebraic equations. Note also that, due to the
structure of the differential equation in (\ref{20}), for
obtaining the correct expressions for the coefficients, we have to
start with a polynomial in (\ref{21}) with terms up to $x^{15/2}$,
rather than $x^{9/2}$.

\begin{figure}
\begin{center}
\epsfysize=7cm \epsfxsize=11cm \epsffile{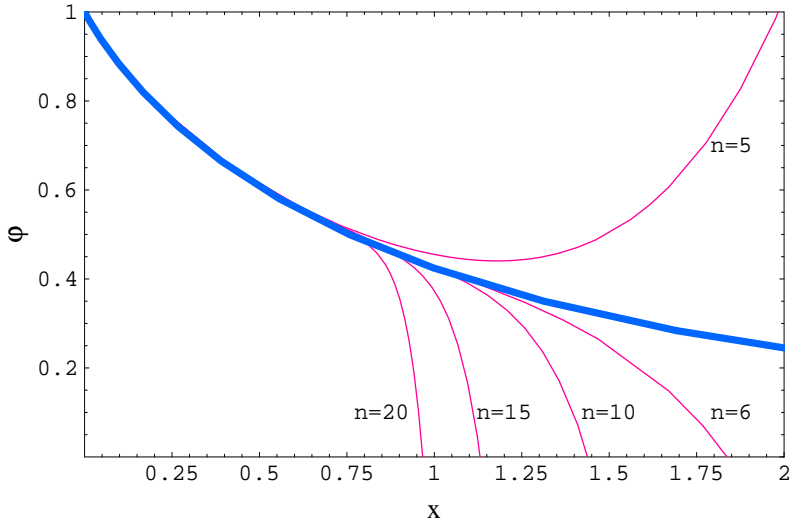} \caption{The
Thomas-Fermi function $\varphi(x)$ (thick curve) and its Frobenius
polynomial approximations for small $x$. The labels
$n=5,6,10,15,20$ refer to the maximum power considered in the
expansion in Eq. (\ref{21}).} \label{fig1}
\end{center}
\end{figure}

\subsection{Exact and semi-analytical results}
The work described above on the polynomial approximation,
performed by Majorana, was only the first step towards an exact or
semi-analytical solution to the Fermi equation. We will indulge
here on an anecdote reported by Rasetti \cite{[FNM]}, Segr\`{e}
\cite{[Segre]} and Amaldi \cite{[Amaldi]}. According to the last
author, ``Fermi gave a broad outline of the model and showed some
reprints of his recent works on the subject to Majorana, in
particular the table showing the numerical values of the so-called
Fermi universal potential. Majorana listened with interest and,
after having asked for some explanations, left without giving any
indication of his thoughts or intentions. The next day, towards
the end of the morning, he again came into Fermi's office and
asked him without more ado to draw him the table which he had seen
for few moments the day before. Holding this table in his hand, he
took from his pocket a piece of paper on which he had worked out a
similar table at home in the last twenty-four hours, transforming,
as far as Segr\`{e} remembers, the second-order Thomas-Fermi
non-linear differential equation into a Riccati equation, which he
had then integrated numerically."

 The whole work performed by Majorana on the solution of the Fermi
equation, is contained in some spare sheets conserved at the Domus
Galilaeana in Pisa, and diligently reported by the author himself
in his notebooks \cite{[Volumetti]}. The reduction of the Fermi
equation to an Abel equation (rather than a Riccati one, as
confused by Segr\`{e}) proceeds as follows. Let's adopt a change
of variables, from $(x, \varphi)$ to $(t, u)$, where the formula
relating the two sets of variables has to be determined in order
to satisfy, if possible, both the boundary conditions (\ref{12}).
The function $\varphi$ in Eq. (\ref{16}) has the correct behavior
for large $x$, but the wrong one near $x=0$, so that we could
modify the functional form of $\varphi$ to take into account the
first condition in (\ref{12}). An obvious modification is
$\varphi=(144/x^3)f(x)$, with $f(x)$ a suitable function which
vanishes for $x\rightarrow 0$ in order to account for $\varphi
(x=0) = 1$. The simplest choice for $f(x)$ is a polynomial in the
novel variable $t$, as it was also considered later, in a similar
way, by Sommerfeld \cite{[Sommerfeld]}. The Majorana choice is as
follows:
\begin{equation}
\varphi (x)= \frac{144}{x^3}(1-t)^2 ,
\label{23}
\end{equation}
with $t\rightarrow 1$ as $ x\rightarrow 0$ \footnote{The
explanation given here of the method pursued by Majorana has been
inferred from the unpublished papers left by the author. However,
differently from the vast majority of arguments treated by
Majorana in his notebooks, no clear explanation of what the author
does is explicitly reported.}. From Eq. (\ref{23}) we can then
obtain the first relation linking $t$ to $x, \varphi$. The second
one, involving the dependent variable $u$, is that typical of
homogeneous differential equations (like the Fermi equation) for
reducing the order of the equation, i.e. exponentiation with an
integral of $u(t)$. The transformation relations are thus:
\begin{equation}
\begin{array}{rcl}
t &=& \displaystyle 1- \frac{1}{12}\sqrt{x^3\varphi},
\\ & & \\
\varphi &=& e^{\int_1^t u(t) dt}\, .
\end{array}
\label{24}
\end{equation}
Substitution into Eq. (\ref{11}) leads to an Abel equation for
$u(t)$,
\begin{equation}
\frac{du}{dt}= \alpha (t) + \beta (t)\, u + \gamma (t)\, u^2 +
\delta (t) \, u^3 , \label{25}
\end{equation}
with
\begin{equation}
\begin{array}{rcl}
\alpha (t) &=& \displaystyle \frac{16}{3(1-t)},
\\ & & \\
\beta (t) &=& \displaystyle 8 +  \frac{1}{3(1-t)},
\\ & & \\
\gamma (t) &=& \displaystyle \frac{7}{3} - 4t,
\\ & & \\
\delta (t) &=& \displaystyle -\frac{2}{3}t(1 - t) .
\end{array}
\label{26}
\end{equation}
Note that both the boundary conditions in (\ref{12}) are
automatically verified by the relations (\ref{24}). We have
reported the derivation of the Abel equation (\ref{25}) mainly for
historical reasons (see also the next section); the precise
numerical values for the Fermi function $\varphi (x)$ were
obtained by Majorana by solving a different first-order
differential equation \cite{[Esposito]}. It is also remarkable
that none of the Majorana's colleagues and friends was aware of
this, which however is well documented in the notebooks
\cite{[Volumetti]} and in some other unpublished papers. Instead
of Eq. (\ref{23}), Majorana chooses $\varphi (x) $ of the form
\begin{equation}
\varphi (x)= \frac{144}{x^3} \, t^{6} . \label{27}
\end{equation}
Now the point with $x=0$ corresponds to $t=0$. In order to obtain again a
first order differential equation for $u(t)$, the transformation equation
for the variable $u$ involves $\varphi$ and its first derivative. Majorana
then introduced the following formulas:
\begin{equation}
\begin{array}{rcl}
t &=& 144^{-1/6} \, x^{1/2} \, \varphi^{1/6}, \nonumber \\ & & \\
u &=& \displaystyle -\left(\frac{16}{3}\right)^{1/3}
\varphi^{-4/3} \varphi^\prime \, \, .
\end{array}
\label{28}
\end{equation}
By taking the $t$-derivative of the last equation in (\ref{28})
and inserting Eq. (\ref{11}) in it, one gets:
\begin{equation}
\frac{du}{dt}= -
\left(\frac{16}{3}\right)^{1/3}\dot{x}\varphi^{-4/3}\left[-\frac{4}{3}
\frac{\varphi^{\prime 2}}{\varphi} +
\frac{\varphi^{3/2}}{x^{1/2}}\right] . \label{29}
\end{equation}
By using Eqs. (\ref{28}) to eliminate $x^{1/2}$ and
$\varphi^{\prime 2}$, the following equation results:
\begin{equation}
\frac{du}{dt}= \left(\frac{4}{9}\right)^{1/3}\frac{tu^2
-1}{t}\dot{x}\varphi^{1/3} . \label{30}
\end{equation}
Now the quantity $\dot{x}\varphi^{1/3}$ can be expressed in terms
of $t$ and $u$ by making use again of the first equation in
(\ref{28}) (and its $t$-derivative). After some algebra, the final
result for the differential equation for $u(t)$ is:
\begin{equation}
\frac{du}{dt}= 8 \, \frac{tu^2 -1}{1-t^2u} . \label{31}
\end{equation}
The obtained equation is again non-linear but, differently from
the original Fermi equation (\ref{11}), it is first-order in the
novel variable $t$ and the degree of non-linearity is lower than
that of Eq. (\ref{11}). The boundary conditions for $u(t)$ are
easily taken into account from the second equation in (\ref{28})
and by requiring that for $x\rightarrow\infty$ the Sommerfeld
solution (Eq. (\ref{27}) with $t=1$) be recovered:
\begin{equation}
\begin{array}{rcl}
u(0) &=& \displaystyle - \left(\frac{16}{3}\right)^{1/3}
\varphi^\prime_0 ,
\\ & & \\
u(1) &=&  1 .
\end{array}
\label{32}
\end{equation}
Here we have denoted with $\varphi^\prime _0 =
\varphi^\prime(x=0)$ the initial slope of the Thomas-Fermi
function $\varphi (x)$ which, for a neutral atom, is approximately
equal to $-1.588$.

The solution of Eq. (\ref{31}) was achieved by Majorana in terms
of a series expansion in powers of the variable $\tau = 1-t$:
\begin{equation}
u= a_0 + a_1 \tau + a_2 \tau^2 + a_3 \tau^3 + ...\,\, .
\label{33}
\end{equation}
Substitution of Eq. (\ref{33}) (with the conditions in Eq.
(\ref{32})) into Eq. (\ref{31}) results into an iterative formula
for the coefficients $a_n$ (for details see Ref.
\cite{[Esposito]}). It is remarkable that the series expansion in
Eq. (\ref{33}) is uniformly convergent in the interval $[0,1]$ for
$\tau$, since the series $\sum_{n=0}^\infty a_n$ of the
coefficients converges to a finite value determined by the initial
slope $\varphi^\prime_0$. In fact, by setting $\tau = 1$ ($t=0$)
in Eq. (\ref{33}) we have from Eq. (\ref{32}):
\begin{equation}
\sum_{n=0}^\infty a_n =  -
\left(\frac{16}{3}\right)\varphi^\prime_0 \label{34}
\end{equation}
Majorana was aware \cite{[Volumetti]} of the fact that the series
in Eq. (\ref{33}) exhibits geometrical convergence with $a_n/
a_{n-1}\sim 4/5$ for $n\rightarrow\infty$.

Given the function $u(t)$, we now have to look for the
Thomas-Fermi function $\varphi(x)$. This was obtained in a
parametric form $x=x(t)$, $\varphi = \varphi (t)$ in terms of the
parameter $t$ already introduced in Eq. (\ref{28}), and by writing
$\varphi (t)$ as
\begin{equation}
\varphi (t) = e^{\int_0^t w(t) dt}
\label{35}
\end{equation}
(with this choice, $\varphi (t=0) =1$ and the first condition in
(\ref{12}) is automatically satisfied). Here $w(t)$ is an
auxiliary function which is determined in terms of $u(t)$ by
substituting Eq. (\ref{35}) into Eq. (\ref{28}). As a result, the
parametric solution of Eq. (\ref{11}), with boundary conditions
(\ref{12}), takes the form:
\begin{equation}
\begin{array}{rcl}
x(t)& = & 144^{1/3} \, t^2 \, e^{2{\cal I}(t)}
\\ & & \\
\varphi (t)& = & e^{-6 \, {\cal I}(t)}
\end{array}
\label{36}
\end{equation}
with
\begin{equation}
{\cal I}(t) = \int_0^t \frac{ut}{1-t^2u} \, dt \label{37}
\end{equation}
Remarkably, the Majorana solution of the Thomas-Fermi equation is
obtained with only one quadrature and gives easily obtainable
numerical values for the electrostatic potential inside atoms. By
taking into account only $10$ terms in the series expansion for
$u(t)$, such numerical values approximate the values of the exact
solution of the Thomas-Fermi equation with a relative error of the
order of $0.1\% $ \cite{[Esposito]}.

\begin{figure}
\begin{center}
\epsfysize=7cm \epsfxsize=11cm \epsffile{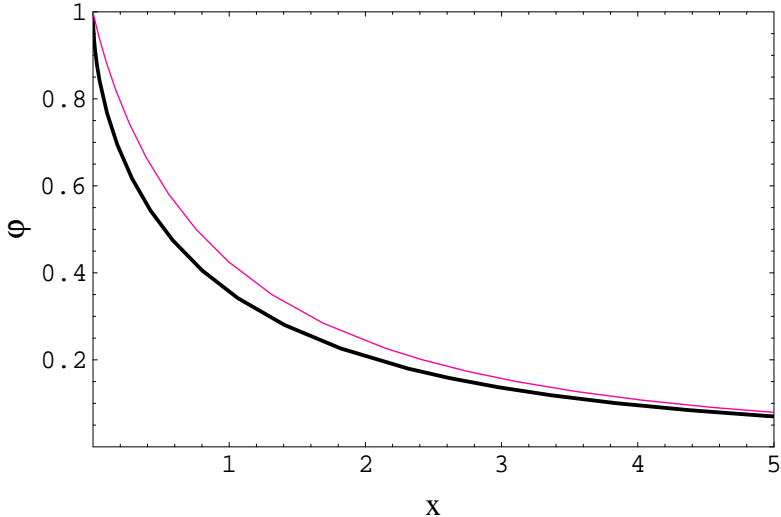} \caption{The
Thomas-Fermi function $\varphi (x)$ and the Majorana approximation
of it. The thin (upper) line refers to the exact (numerical)
solution of Eq. (\ref{11}) while the thick (lower) one corresponds
to the parametric solution obtained from Eqs.
(\ref{36})-(\ref{37}).} \label{fig2}
\end{center}
\end{figure}

\subsection{Numerical tables}
It is instructive to give a look at the numerical results for the
Thomas-Fermi function $\varphi (x)$ obtained by several authors
with different methods. As already pointed out, Thomas firstly
reported numerical values for the potential inside atoms. Since
his approach involved a differential equation equivalent to the
Fermi equation but with a different mathematical structure (see
\cite{[Thomas]}), which does not make direct use of the universal
function $\varphi (x)$, the comparison between the Thomas
numerical results and those obtained by Fermi and others will not
be considered here.

It is a matter of fact that many atomic physicists in the $'30$s
used the Fermi table in their computations. As early as at the end
of $1927$, according to Segr\`{e} \cite{[Segre]} and Rasetti
\cite{[FNM]}, Fermi obtained the values of the function $\varphi
(x)$ by means of successive approximation in Eq. (\ref{11}),
during approximatively one week of numerical work, by using a
small hand calculator (a Brunsviga one). The table, however, was
published later in a German paper \cite{[FN49]}; we reproduce in
Table \ref{t1} some of these results.\footnote{In Table \ref{t1}
we report only the results obtained by Fermi, Majorana and Miranda
corresponding to common values of $x$ considered by all three
authors.} As pointed out above, Majorana checked the Fermi results
by using probably the parametric solution in Eqs. (\ref{36}). Note
that only the integration in Eq. (\ref{37}) requires numerical
evaluation. The results obtained by Majorana, during
approximatively one night of numerical work, are reported in his
``Volumetti" \cite{[Volumetti]} and are reproduced here in Table
$1$.

A rapid look at this table shows a satisfactory agreement between
the Fermi numerical approach and the Majorana method. It is even
remarkable that, for the first points, Majorana obtained also the
values for the derivative of the function $\varphi (x)$ (see
\cite{[Volumetti]}). Subsequent more accurate numerical
evaluations of the solution of the Thomas-Fermi equation were
performed by Miranda in $1934$ \cite{[Miranda]}, who also gave a
solid mathematical framework to the numerical integration of the
Fermi equation (see the next section). By using refined
approximation procedures to a finite variation equation
corresponding to the Fermi differential equation, he obtained
numerical values for $\varphi (x)$ (and $\varphi^\prime(x)$) which
are accurate up to the fifth significant digit for small $x$
(Fermi and Majorana results are accurate ``only" up to the third
significant digit in the same $x$ interval). Some results
(corresponding to an initial slope of $-1.5880464$) are reproduced
here in Table \ref{t1} for comparison.
\begin{table}
\caption{Numerical values of the function $\varphi (x)$ taken from
the Fermi, Majorana and Miranda papers.} \[
\]\centering
\begin{tabular}{ll}
\hline $x$ & $\varphi_{\rm Fer}$
\\ \hline
0 & 1
\\
0.1 & 0.882
\\
0.2 &  0.793
\\
0.3 &   0.721
\\
0.4 &   0.660
\\
0.5 &   0.607
\\
0.6 &   0.562
\\
0.7 &  0.521
\\
0.8 &  0.485
\\
0.9 & 0.453
\\
1 &  0.425
\\
1.2 &  0.375
\\
1.4 &  0.333
\\
2 &  0.244
\\
3 &  0.157
\\
\hline
\end{tabular} $\!\!\!\!\!$
\begin{tabular}{l}
\hline $\varphi_{\rm Maj}$
\\ \hline
1
\\
0.882
\\
0.793
\\
0.721
\\
0.660
\\
0.607
\\
0.561
\\
0.521
\\
0.486
\\
0.453
\\
0.424
\\
0.374
\\
0.333
\\
0.243
\\
0.157
\\
\hline
\end{tabular} $\!\!\!\!\!$
\begin{tabular}{l}
\hline $\varphi_{\rm Mir}$
\\ \hline
1
\\
0.88170
\\
0.79307
\\
0.72065
\\
0.65955
\\
0.607
\\
0.56118
\\
0.52081
\\
0.48495
\\
0.45288
\\
0.42403
\\
0.37427
\\
0.33294
\\
0.24306
\\
0.15675
\\
\hline
\end{tabular}
~~~~~
\begin{tabular}{ll}
\hline $x$ & $\varphi_{\rm Fer}$
\\ \hline
 4 & 0.108
\\
 5 &  0.079
\\
 6 & 0.059
\\
 7 & 0.046
\\
 8 &  0.037
\\
 9 &  0.029
\\
10 & 0.024
\\
20 &  0.0056
\\
30 &  0.0022
\\
40 &  0.0011
\\
50 &  0.00061
\\
60 &  0.00039
\\
80 & 0.00018
\\
100 &  0.0001
\\
${}$ &  ${}$
\\
\hline
\end{tabular} $\!\!\!\!\!$
\begin{tabular}{l}
\hline $\varphi_{\rm Maj}$
\\ \hline
0.108
\\
0.079
\\
0.059
\\
0.046
\\
0.036
\\
0.029
\\
0.024
\\
0.0056
\\
0.0022
\\
0.0011
\\
0.0006
\\
0.0004
\\
0.0002
\\
0.0001
\\
${}$
\\
\hline
\end{tabular} $\!\!\!\!\!$
\begin{tabular}{l}
\hline $\varphi_{\rm Mir}$
\\ \hline
0.1086
\\
0.0798
\\
0.0599
\\
0.0469
\\
0.038
\\
0.030
\\
0.025
\\
0.0063
\\
0.0024
\\
0.0012
\\
0.00068
\\
0.00042
\\
0.00019
\\
0.0001
\\
${}$
\\
\hline
\end{tabular} \label{t1}
\end{table}
A remarkable agreement between the results given by the three
authors, which used completely different techniques, can be
clearly deduced.

\subsection{Mathematical properties}
The non-linear second-order differential equation (\ref{11}) has
received much attention by physicist from its discovery until now,
mainly due to the important physical model underlying it, which is
not limited to atoms \cite{[nuclear]}. On the mathematical side,
some formal properties of the solutions of the Thomas-Fermi
equation have been studied as well. We will give here a brief
account of the results achieved in the literature, excluding from
our discussion those corresponding to asymptotic expressions and
theorems on numerical approximations, which have been already
considered above.

From the mathematical point of view, the starting most important
result for a non-linear differential equation is the theorem
establishing the existence and uniqueness of its solutions. For
the case considered here, we point out that the physically
interesting solutions of Eq. (\ref{11}) are those satisfying the
boundary conditions (\ref{12}) (or similar ones for electrically
charged ions).

The studies performed by Majorana and described above, especially
those aimed to transform the Thomas-Fermi equation into an Abel
equation, seem to leave little space to speculations in this
direction. In fact, the theorem of existence and uniqueness for
$\varphi (x)$ directly follows from that holding for the Abel
equation (\ref{25}), allowing the integrability of its solution
$u(t)$ which is required in Eqs. (\ref{24}).

However, the Majorana work on the Thomas-Fermi equation was
practically unknown to everybody (until recent times
\cite{[Volumetti]}, \cite{[Esposito]}), and some other subsequent
papers are usually quoted in the literature. The theorem of
existence and uniqueness for Eq. (\ref{11}) with conditions
(\ref{12}) was clearly stated in $1929$ by two Italian
mathematicians, Mambriani \cite{[Mambriani]} and Scorza-Dragoni
\cite{[Scorza]}. Looking at the analytical properties of the
function appearing in the R.H.S. of Eq. (\ref{11}) and using
standard methods holding for ordinary differential equations, they
showed that an infinite number of integral curves of Eq.
(\ref{11}) pass through a given point $(x_0, \varphi_0)$ of the
first quadrant of the $x-\varphi$ plane, only one of which
$\varphi^*(x)$ being always decreasing and approaching the
$x$-axis. This solution corresponds to a ``critical" value of the
initial slope $\varphi^\prime_0$, given by $-1.588$. The other
solutions with an initial slope greater than the critical value
lie above $\varphi^*(x)$ and diverge for diverging $x$, while
those with an initial slope lower than the critical value lie
below $\varphi^*(x)$ and monotonically decrease for increasing $x$
as far as they reach the $x-$axis.

Mambriani and Scorza-Dragoni also gave a generalization of the
theorem above applicable to a given class of differential
equations \cite{[Mambriani]}, \cite{[Scorza]}. It is also
interesting to point out how several mathematicians participated
in the debate originated by the discussion of the Thomas-Fermi
differential equation, including Caccioppoli as quoted in
\cite{[Scorza]}.

Some other mathematical questions underlying the statistical model
introduced by Thomas and Fermi and the solutions of the
corresponding equation were then tackled only later in the
$1940$'s \cite{[Feynman]}. In particular the attention drifted
towards variational methods applied to a ``Thomas-Fermi energy
functional", involving the screened potential $\varphi(x)$, in
order to obtain the ground state energy for the relevant atoms
(and molecules). A renewed interest started in $1969$ with the
paper by Hille \cite{[Hille]}, who studied analytically a number
of mathematical aspects of the Thomas-Fermi equation for the
atomic case. Generalization to the molecular case was subsequently
analyzed by Lieb and Simon in $1977$ \cite{[Liebsimon]}, proving
the existence and uniqueness of the corresponding Thomas-Fermi
function.

We address the interested reader to the quoted literature for the
details on this subject, which is beyond the aims of the present
paper.

\section{First applications of the statistical method}
The first practical applications of the Thomas-Fermi model of
atoms were developed by Fermi himself and his collaborators in
Rome (mainly Rasetti \cite{[RasettiM]}, Gentile and Majorana
\cite{[Gentile]}). According to Rasetti \cite{[FNM]}, ``this work
had been the main activity of the Rome Institute in 1928". As
outlined in Sect. 2, the electrostatic potential inside an atom of
atomic number $Z$, at a distance $r$ from the nucleus, can be cast
in the form:
\begin{equation}
V(r)=\frac{Ze}{r}\varphi \left(\frac{r}{\mu}\right), \label{40}
\end{equation}
where $\varphi $ is the Fermi function discussed above and $\mu$
is given in Eq. (\ref{10}). Based on this, Fermi \cite{[FN44]}
calculated the number of electrons in an atom with given values of
the orbital angular momentum as function of the atomic number $Z$
\footnote{It is interesting to compare the deduction by Fermi in
\cite{[FN44]} with the one made by Majorana in Sect. 10 of
Volumetto II, reported in \cite{[Volumetti]}.}, thus succeeding to
give an account of the appearance of the elements in the periodic
table. Of course it was readily realized that, since the potential
$V(r)$ was determined by means of statistical arguments, only
average properties of the periodic system can be explained in
terms of the Thomas-Fermi model. Peculiarities of the electronic
structure, underlying particular properties of the elements,
cannot be accounted for by a statistical method. Nevertheless, as
noted by Fermi himself, the agreement with the experimental
observations is satisfactory.

The next step was to evaluate the energy levels of the quantum
states of all (heavy) elements. This was done by Fermi in
\cite{[FN45]} for the S-levels; in particular he calculated the
Rydberg correction for the S-levels of any element from the
approximate solution of the Schr\"odinger equation for an
s-electron with zero energy. The energy levels corresponding to
the angular momentum $l=3, m=2$ (M-levels) for the X-ray spectrum
were, instead, considered by Rasetti \cite{[RasettiM]}. Again the
experimental points fluctuate closely about the predicted values.
A slightly different application of the Thomas-Fermi potential was
carried out by Gentile and Majorana \cite{[Gentile]}, who
calculated the doublet separation due to spin for optical and
X-ray levels of some elements, applying the Dirac theory. Moreover
they also evaluated the ratio of the intensities of the absorption
s-p lines in the alkali spectra. Some further applications were
performed by the Fermi group during $1928$, mainly devoted to
explain the properties of the rare earths and the electron
affinity of the halogens, which were discussed in a restricted
conference in Leiprig under the chairmanship of P. Debye
\cite{[FN49]}.

Several generalization of the Thomas-Fermi method for atoms were
proposed as early as in $1928$ by Majorana, but they were
considered by the physics community, ignoring the Majorana
unpublished works, only many years later (see, for example, the
review in \cite{[Spuch]}).

Indeed, in Sect. 16 of Volumetto II \cite{[Volumetti]}, Majorana
studied the problem of an atom in a weak external electric field
$E$, i.e. atomic polarizability, and obtained an expression for
the electric dipole moment for a (neutral or arbitrarily ionized)
atom.

Furthermore, he also started to consider the application of the
statistical method to molecules, rather than single atoms,
studying the case of a diatomic molecule with identical nuclei
(see Sect. 12 of Volumetto II \cite{[Volumetti]}). The effective
potential in the molecule was cast in the form:

\begin{equation}
V= V_1 +V_2 -\alpha \, \frac{2V_1V_2}{V_1 + V_2}, \label{41}
\end{equation}

\noindent $V_1$ and $V_2$ being the potentials generated by each
of the two atoms. The function $\alpha$ must obey the differential
equation for $V$,
\begin{equation}
\nabla^2 V =-kV^{3/2} \label{42}
\end{equation}
($k$ is a suitable constant), with appropriate boundary
conditions, discussed in \cite{[Volumetti]}. Majorana also gave a
general method to determine $V$ when the equipotential surfaces
are approximately known (see Sect. 12 of Volumetto III
\cite{[Volumetti]}). In fact, writing the approximate expression
for the equipotential surfaces, as functions of a parameter $p$,
as
\begin{equation}
f(x,y,z) =p, \label{43}
\end{equation}
he deduced a thorough equation from which it is possible to
determine $V(\rho)$, when the boundary conditions are assigned.
The particular case of a diatomic molecule with identical nuclei
was, again, considered by Majorana using elliptic coordinates in
order to illustrate his original method \cite{[Volumetti]}.

\section{Conclusions}
In this paper we have depicted the genesis and the first
developments of the statistical model of atoms introduced by
Thomas and Fermi in 1926-27. Far from being complete, our account
has focused on the results achieved by Fermi and his group in
Rome, as given evidence by many articles published in widespread
journals. We have also pointed out the practically unknown
contribution to the model given by Majorana, who was introduced to
the subject by Fermi himself. One of the major results reached by
Majorana as early as in the beginning of 1928 is its solution (or,
rather, methods of solutions) of the Thomas-Fermi equation. This
plays an important role in the rapid and accurate determination of
the effective electrostatic potential in atoms, required in
physical applications, as well as in studying the mathematical
properties of the differential equation itself, anticipating later
(Mambriani, Scorza-Dragoni, Miranda 1929-1934) and much later
(Hille 1969) achievements. Majorana works on these topics (as well
as in many other ones) is contained in his notebooks
\cite{[Volumetti]} and was not published by the author: it is
remained unknown until recent times \cite{[Volumetti]}. A brief
account of the statistical model of atoms, which is widely known,
has been reported in Sect. 2, and follows quite closely the Fermi
approach. \footnote{See also Sects. 8, 9 and 10 of Volumetto II
\cite{[Volumetti]} by Majorana, where the author practically
parallel the first three papers by Fermi \cite{[FN43]},
\cite{[FN44]}, \cite{[FN45]} on this subject.} Wide room has been
made to the solution of the Thomas-Fermi equation in Sect. 3; the
interested reader, looking for more technical details, can benefit
from a reading of papers \cite{[Esposito]}, \cite{[EspositoOde]}.
Early applications of the Thomas-Fermi model, performed by the
Fermi group (as well as some other people), essentially dealt with
atomic spectroscopy, and have been discussed above in Sect. 4.
Moreover we have also highlighted some completely novel (for that
time) applications of the statistical model by Majorana, who
employed it for studying atoms in external fields (atomic
polarizability) and molecules.

From what discussed here, it is then evident the considerable
contribution given by Majorana even in the achievement of the
Thomas-Fermi model, anticipating, in many respects, some important
results reached later by leading specialists.

\vspace{2cm} \noindent {\Large \bf Acknowledgments}

\vspace{1truecm}

\noindent The authors are indebted with Prof. E. Recami; Dr. E.
Majorana jr. and Prof. A. Drago for fruitful discussions.

\end{document}